\documentclass[12pt]{article}

\usepackage{cite}

\def\beq{\begin{eqnarray}}   \def\eeq{\end{eqnarray}}

\newcommand{\ntwo}{${\cal N}\!\!=\!2\,$}
\newcommand{\none}{${\cal N}\!\!=\!1\,$}

\begin{document}
\begin{titlepage}
\begin{flushright}
TPI-MINN-40/01  \\
UMN-TH-2022\\
ITEP-TH-44-01\\
hep-th/0108153\\
\end{flushright}

\vfil

\begin{center}
{\Large\bf Counting Supershort Supermultiplets}
\end{center}

\begin{center}
{\bf A.~Losev$^{\,a,b}$,  M.~Shifman$^{\,a}$, and A.~Vainshtein$^{\,a}$}
 \end{center}
\vspace{0.1cm}

\begin{center}
$^a$  {\em Theoretical Physics Institute, University of Minnesota, \\
116 Church St SE, Minneapolis, MN 55455, USA}\\
~\\
$^b$ {\em Institute of Theoretical and Experimental Physics,\\
B. Cheremushkinskaya 25, Moscow
117259, Russia}\,\footnote{Permanent address}
 \end{center}

\vspace{0.9cm}

\begin{abstract}
We consider multiplet shortening for BPS solitons in ${\cal N}\!\!=\!1\,$
two-dimensional models. Examples of the single-state multiplets
were established previously in ${\cal N}\!\!=\!1\,$ Landau-Ginzburg models. 
The shortening comes at a price of loosing the fermion parity $(-1)^F$ due 
to boundary effects. This implies the disappearance of the boson-fermion 
classification resulting in abnormal statistics. To count such short 
multiplets we introduce a new index. We consider the phenomenon of shortening 
in a broad class of hybrid models which extend the Landau-Ginzburg models 
to include a nonflat metric on the target space. Our index turns out to be 
related to the index of the Dirac operator on the soliton moduli space. 
The latter vanishes in most cases implying the absence of shortening. 
We also generalize the anomaly in the central charge to take into account 
the target space metric.

\end{abstract}

\vspace{0.5cm}

\end{titlepage}
   
\newpage

\section{Introduction}

Supersymmetry (SUSY) unites bosons and fermions  into multiplets of degenerate 
states. It seems to imply that the minimal SUSY multiplet contains at least 
two states: one bosonic and one fermionic. Nevertheless, an example of the
supermultiplet consisting of only one state is provided by a BPS soliton in
 \none two-dimensi\-onal theories (see \cite{1d}). 
It is clear that this soliton is neither boson nor fermion --- the fermion 
parity $(-1)^F$ becomes ill-defined in the soliton sector.  The phenomenon 
is similar in many respects  to the appearance of fractional charges for 
the 2D solitons~\cite{JR}.  

The supersymmetry
algebra in the \none two-dimensional theories (often denoted as
${\cal N}\!=\!\{1,1\}$) is formed by  two real supercharges $Q_{\alpha }$, 
($\alpha =1,2$), the energy-momentum vector $P_\mu $, ($\mu
=1,2$), and  the central charge 
${\cal Z}$,
\begin{equation}
\{ Q_\alpha\,,  {\bar Q}_\beta \}=2\left(\gamma^\mu 
P_\mu +\gamma^5 {\cal Z}\right)_{\alpha\beta}\, ,
\qquad 
\left[Q_\alpha, P_\mu\right]=\left[Q_\alpha, {\cal Z}\right]=0\,.
\label{alg}
\end{equation}
Here ${\bar Q}_\beta=Q_\alpha (\gamma^0)_{\alpha\beta}$ and
$
\gamma^0\!=\!\sigma_2\,, \;\gamma^1\!=\! i\sigma_3\,,\;
\gamma^5\!=\! i\gamma^0\gamma^1\!=\! -i\sigma_1
$
 are purely imaginary two-by-two
matrices.
The central charge is nonvanishing in the soliton sector, and we define  ${\cal
Z}\!>\!0$ for the soliton (${\cal Z}\!<\!0$ for the antisoliton). 

In the rest frame of the soliton, $P_\mu =(M,0)$,  the
algebra (\ref{alg}) takes the form 
\begin{equation}
Q_1^2=M+{\cal Z}\,,\qquad Q_2^2=M-{\cal Z}\,,\qquad
\{ Q_1\,,  {Q}_2 \}=0\,.
\label{algrest}
\end{equation}
For the BPS soliton $Q_2|\,{\rm sol}\rangle=0$ implying that its  mass $M$ is 
equal to the central charge
${\cal Z}$. Then it is clear that $Q_1$ equal to $\sqrt{2{\cal Z}}$ 
(or $-\sqrt{2{\cal Z}}$) leads to the  irreducible representation of the 
superalgebra which is {\em one-dimensional}.  

As was mentioned above in the one-dimensional representation 
the fermion parity $(-1)^F$ is not defined. 
Usually, it is implied, however, that  $(-1)^F$ does exist. Indeed, 
microscopically we start with the local field theories where classification 
of the
fields as bosonic and fermionic is explicit.  For supercharges 
the fermion parity is $-1$ which makes the representation 
reducible --- consisting of two one-dimensional representations 
\footnote{Note, 
that in the same way the one-dimensional  representation appears  for the
massless particles in 2+1 dimensions --- the superalgebra there has the same 
form
(\ref{alg}) with identification of
${\cal Z}$ and $\gamma^5$ as an extra component of $P_\mu $ and
$\gamma^\mu$. Maintaining $(-1)^F$ makes the representation two-dimensional
and reducible, see e.g. \cite{BdW}.} .

How come that in the soliton sector $(-1)^F$ becomes
ill-defined? This happens due to boundary effects.  A technical signal  is
the existence of only one normalizable fermion zero mode on the BPS soliton.
Another fermion zero mode (concentrated at the boundary)  would appear if a
finite box with  proper boundary conditions were introduced. For physical
measurements made far away from the boundary the fermion parity $(-1)^F$
is lost, and the one-dimensional multiplet becomes a physical reality.
 It is similar to the effect of the fractional charge: the total charge
which includes boundaries stays integral but it is irrelevant for local 
experiments. 

Once the boson-fermion classification for solitons is lost, it clearly implies
 certain abnormalities in the soliton statistics~\footnote{The 
abnormal statistics were extensively discussed in the literature, see e.g.
\cite{Wilczek,Clif}.}. In particular,
the statistical counting of the multiplicity of states 
\cite{Tsvelik,Zamol,WITTENlecture} (the multiplicity of states $K$ is defined
through the entropy  per soliton of a  gas of noninteracting solitons, 
$S=\ln K$) yields $K=\sqrt{2}$. 
This unusual result was recently confirmed  by Fendley and Saleur
\cite{FendleySaleur}.  In their stimulating work the Gross-Neveu model with
$N$ fermions was treated within  the thermodynamical Bethe {\em ansatz}. For
$N\!=\!3$ the model coincides  with the SUSY sine-Gordon model earlier
considered by Tsvelik \cite{Tsvelik} within a different method. 

In this paper we address a few  issues related
to BPS solitons in \none theories
in 1+1 dimensions.
We consider a class of \none
hybrid models 
 where along with the superpotential ${\cal W}(\phi)$
there is a nonflat metric $g_{ab}(\phi)$ of the target space for the 
fields $\phi^a$
(hybrids between the sigma models and Landau-Ginzburg models).
Within these models we identify those in which
the multiplet shortening does {\em not} take place.
In fact, the existence of the
BPS solitons belonging to one-dimensional supermultiplets
turns out to be  quite a  rare occasion.

We introduce a new index which counts such short multiplets. Let us remind 
that 
the first SUSY index, ${\rm Tr}\, (-1)^F$, was introduced by Witten
twenty years ago \cite{WittenIndex} to count the number of supersymmetric 
vacua.
About ten years ago, Cecotti, Fendley, Intriligator and Vafa
 introduced  \cite{CFIV} another index,
 Tr$[F\,(-1)^F]$, counting the number of short multiplets in \ntwo
theories in two dimensions. No index counting single-state
multiplets in \none theories 
in two dimensions (two supercharges) was known. This is probably not
surprising,  since it was always assumed that $(-1)^F$ does exist.  

Our task is to find
such index for \none.
We will show that the index is $\left\{{\rm Tr}\,Q_1\right\}^2\!/2{\cal
Z}$ --- it vanishes for long multiplets and is equal to
1 for one-dimensional multiplets.
If the value of this index 
  does not vanish
in the given \none theory, 
short multiplets {\em do exist} with necessity.

Another issue we address is the 
 generalization of the anomaly in the central
charge found previously in the Landau-Ginzburg models \cite{SVV}, where the
target space metric is flat, to the hybrid models with
 a nonflat metric. Our result for the anomaly in this
case is a straightforward extension of \cite{SVV} and can be formulated as 
a substitution 
\begin{equation}
{\cal W}(\phi) \longrightarrow \widetilde {\cal W}(\phi )={\cal
W}(\phi)+\frac{1}{4\pi}\,\nabla^a
\nabla_a {\cal W}(\phi)
\label{anoW}
\end{equation}
for the superpotential. The quantum anomaly is represented by term with 
the covariant Laplacian on
the target space, $\nabla^a \nabla_a\equiv g^{ab}\nabla_a\nabla_b\,$.
The anomaly corrected superpotential enters  into the energy-momentum tensor,
the supercharges and the central charge. In particular, the operator of the 
central charge  becomes
\begin{equation}
{\cal Z} = 
\widetilde{\cal W} (\phi (z\to \infty))-\widetilde{\cal W} (\phi (z\to 
-\infty))
\,.
\end{equation}

A more detailed discussion of the above issues and our results will be given
elsewhere \cite{LSV1}.

 \section{The model}

A generic hybrid model in two dimensions, with
two supercharges contains $k$ real bosonic fields $\phi ^a$ ($a=1,\ldots,k$)
and the same number of the real two-component fermionic fields $\psi_\alpha^a$
($\alpha =1,2$). The model is characterized 
by a metric $g_{ab}(\phi )$  and a 
superpotential ${\cal W}(\phi )$  defined on the
target space for which $\phi ^a$ serve as coordinates. We assume the target space
to be an arbitrary Riemann manifold ${\cal T}$ and  ${\cal W}(\phi )$ an
arbitrary function which has more than one critical point, i.e. points where
$\partial_a{\cal W}(\phi )=0$, on  ${\cal T}$. 

The general form of the Lagrangian is (for a review see
Ref.~\cite{quantmat})
\begin{eqnarray}
{\cal L}&=&\!\frac{1}{2}\,g_{ab}
\left[\,\partial_\mu\phi^a \,
\partial^\mu\phi^b + \bar\psi^a \,i\gamma^\mu 
{\cal D}_\mu \psi^b +F^a F^b\right]
+ \frac{1}{12} \, R_{abcd}\,(\bar\psi^a \psi^c)
(\bar\psi^b \psi^d)\nonumber\\[1mm]
&+&\! F^a\partial_a {\cal W} - \frac1 2 \,
(\nabla_a \partial_b {\cal W})\,\bar\psi^a
\psi^b
\,,
\label{sigmaL}
\end{eqnarray}
where 
 $F^a$ is the auxiliary field, $F^a=-g^{ab}\partial_b{\cal W}$,
and
$\partial_a$ and
$\nabla_a$ denote the conventional 
 and  covariant derivatives on the target space, e.g. for the vector,
$
\nabla_a V_b=\partial_a\, V_b -
\Gamma^c_{ab} \,V_c\,.
$
The covariantized space-time derivative ${\cal D}_\mu$ is
\begin{equation}
{\cal D}_\mu \, \psi^b=\frac{\partial\, \psi^b}{\partial x^\mu} +
\Gamma^b_{cd}\, 
\frac{\partial\, \phi^c}{\partial x^\mu}\, \psi^d\,.
\label{covd}
\end{equation}
Furthermore, 
$\Gamma^b_{c\,d}(\phi)$ and  $R_{abcd}(\phi)$ stand for 
 the Christoffel symbols
and the Riemann tensor on ${\cal T}$. 

The  \none
 supersymmetry of the model is expressed by two
supercharges,
\begin{equation}
Q_\alpha=\int {\rm d} z \,J_{\alpha}^0\,,\qquad
J^\mu=g_{ab}\left(\not \!\!\,\partial \,\phi^a-i \,F^a\right)
\gamma^\mu\,\psi^b \,,
\label{supercur}
\end{equation}
where $J^\mu$ is the conserved supercurrent.
These supercharges form a centrally
extended  ${\cal N}=1$   superalgebra (\ref{alg})
 with the central charge
\begin{equation}
{\cal Z}_0=\int  {\rm d} z\, \partial_z \phi^a 
\,\partial_a {\cal W}\,.
\label{cch}
\end{equation}
The central charge does not vanish for classical 
solutions interpolating between
different vacua of the theory,  $\phi=A$ at $z=-\infty$ and $\phi=B$ at 
$z=\infty$. 
These vacua  correspond to two distinct critical points of the  superpotential 
${\cal W}(\phi)$.  In the classical approximation for the soliton  
interpolating between the critical points $A$ and $B$  the central  charge 
is equal to 
\begin{equation}
{\cal Z}_0={\cal W}(B)-{\cal W}(A)\,.
\label{zclas}
\end{equation}
The BPS soliton satisfies the following equation:
\begin{equation}
\frac{ {\rm d} \phi^a_{\rm sol}}{{\rm d} z}= \, g^{ab}\,\partial_b \,{\cal
W}(\phi_{\rm sol})\,,
\label{AAone}
\end{equation}
where $z$ is the spatial coordinate.

\section{Anomaly}

Above, the subscript 0 marks a bare central charge ${\cal Z}_0$,
which, unlike \ntwo models, gets renormalized
by quantum corrections. The model (\ref{sigmaL})
is renormalizable provided the manifold ${\cal T}$ is symmetric,
it is super-renormalizable when the metric is flat, e.g., 
$g_{ab}=\delta _{ab}\,$.
In Ref.~\cite{SVV} it was shown that
the super-renormalizable Landau-Ginzburg models 
with flat metric possess an anomaly in the central charge,
a superpartner of the trace anomaly in the energy-momentum
tensor. 
Parallelizing the  derivations presented in Sec. III of \cite{SVV}
it is straightforward to include the target space metric.
The result reduces to the substitution (\ref{anoW}) for the superpotential.
The anomalous part given by action of the covariant Laplacian can be written as
\begin{equation}
\frac{1}{4\pi}\nabla^a\nabla_a {\cal W}=
\frac{1}{4\pi}g^{ab}\nabla_a\partial_b {\cal W}\,.
\label{reduc}
\end{equation}

The expression for the central charge becomes
\begin{equation}
{\cal Z} =\widetilde{\cal W}(B)-\widetilde{\cal W}(A)\,,
\label{hyban}
\end{equation}
where the anomaly corrected superpotential $\widetilde {\cal W}$ defined by
(\ref{anoW}) takes place of ${\cal W}$. It is this expression which gives 
the mass of
the  BPS saturated soliton. Note that  while the classical result 
(\ref{zclas}) for the
central charge ${\cal Z}_0$ is  metric independent a dependence on the 
target space
metric emerges through the anomaly.

Omitting details of the derivation
let us only note that
the form of the anomaly in the hybrid models is constrained
by the following considerations: (i) dimension and locality;
(ii) general covariance in the target space; (iii)
 in the limit of  flat metric
it must coincide with the anomaly established in the 
Landau-Ginzburg models \cite{SVV}.

\section{\mathversion{bold}Tr$\,{Q}_{1}$ as an index}

The construction of the BPS representations  
 was discussed in the Introduction.
In the soliton rest frame the centrally extended ${\cal N}=1$
superalgebra takes the form (\ref{algrest}). 
For the BPS soliton $M= {\cal Z}$, and
\begin{equation}
Q_2|\,{\rm sol}\,\rangle = 0\,, 
\label{1dim}
\end{equation}
while
\begin{equation}
Q_1|\,{\rm sol}\,\rangle=\pm \sqrt{2{\cal Z}}\;|\,{\rm sol}\,\rangle\,. 
\label{q2z}
\end{equation}
Equation (\ref{q2z}) implies that the fermion parity  $(-1)^F$ is not defined 
 for the irreducible one-dimensional supermultiplets.

Is there an index in the ${\cal N}=1$ 
soliton problem which would count the BPS
multiplets, i.e., states annihilated by the supercharge $Q_2\,$? 

We assert that 
$\left\{\,{\rm Tr}\,Q_1\right\}^2$ does the job. 
More exactly,
 the  definition of the index 
 is as follows
\begin{equation}
{\rm Ind}_{\,\cal Z}\,(Q_2/Q_1)=\frac{1}{2\,{\cal Z}}\left\{\lim_{\beta \to
\infty}{\rm Tr}
\left[Q_1\,\exp(-\beta\,(Q_2)^2 )\right]\right\}^2\,.
\label{Ind}
\end{equation}
The exponential factor in Eq. (\ref{Ind})
is introduced for the UV regularization. 
The necessity of taking the $\beta \to\infty$ limit is due to continuous 
spectrum
as is explained in \cite{CFIV}.

This index vanishes for non-BPS multiplets. Indeed, 
 as will be explained shortly,
for non-BPS multiplets the fermion parity $(-1)^F$  can be consistently
defined and ${\rm Tr}\,Q_1$ vanishes along with 
${\rm Tr}\, (-1)^F\,$.
For each irreducible BPS representation the index is unity,
$$
{\rm Ind}_{\,\cal Z}\,(Q_2/Q_1)\,[\,{\rm irreducible~BPS}\,] = 1 \,.
$$

If a reducible representation contains a few irreducible BPS multiplets
the index may or may not vanish.  For the vanishing index one can introduce
$(-1)^F$ and small deformations can destroy the BPS saturation. 
Note, that our index is not additive: it is not equal to the sum of the indices 
of the irreducible representations. 
An interesting example of a \none reducible BPS representation is provided by
solitons in \ntwo models. The \ntwo BPS multiplet consists of two \none
multiplets and the index (\ref{Ind}) vanishes.

Thus, $\left\{\,{\rm Tr}\,Q_1\right\}^2\!/2{\cal Z}$ 
 counts the number of the BPS solitons annihilated by  $Q_2$, in the sector
characterized by the value of the central charge ${\cal Z}$. 

The definition (\ref{Ind}) has a technical drawback --- it refers
to the soliton rest frame. It is simple to make it Lorentz invariant,
\begin{equation}
{\rm Ind}_{\,\cal Z}\,(Q_2/Q_1)=\frac{1}{2\,{\cal Z}^2}\left(\,{\rm Tr}\,
\bar Q\right)\! \not \! P \left(\,{\rm Tr}\, Q\right)
\,,
\label{IndL}
\end{equation}
where the trace refers to the Hilbert space but not to the Lorentz indices
of the supercharges $Q_\alpha$ and $\bar Q_\alpha=Q_\beta
(\gamma^0)_{\beta\alpha}$. Here we have omitted the regularizing exponent.

Let us show now that for non-BPS representations the fermion-boson 
classification, based on $(-1)^F$,
is well defined. For non-BPS solitons, with
$M \!>\!{\cal Z}$,  the irreducible representation of the
 algebra
({\ref{algrest}})  is two-dimensional. For instance, one can choose
\begin{equation}
Q_1=\sigma_1\sqrt{M+{\cal Z}}\,,\qquad Q_2=\sigma_2\sqrt{M-{\cal Z}}\,,
\label{two-d}
\end{equation}
where $\sigma_{1,2}$ are the Pauli matrices. 
The boson-fermion classification in this two-dimensional representation is well
known, the operator $(-1)^F$ (anticommuting with $Q_\alpha$) is represented
by $\sigma_3\,$. 

Generically, for non-BPS supermultiplets one can define $(-1)^F$ in terms of 
supercharges as
\begin{equation}
(-1)^F =\frac{\bar Q\,Q}{2\,\sqrt{M^2-{\cal Z}^2}}\,.
\label{gpar}
\end{equation}
It is obvious that once $(-1)^F$ is defined the trace of the operators 
$Q_\alpha$
 connecting  bosonic and fermionic states vanishes. The trace of $(-1)^F$ also 
vanishes since ${\rm Tr}\,\bar Q\,Q=-i \,{\rm Tr}\,[\,Q_1\,,Q_2]=0$.
 
Note, that for non-BPS multiplets not only the fermion parity but also 
$F$ as a generator of a U(1) symmetry  can be defined, namely
\begin{equation}
F=\frac{1}{2}\left[1-(-1)^F \right]\,.
\label{FC}
\end{equation}
Let us emphasize, however,   that unlike  \ntwo
models,  no local current associated with $F$
 exists, and
the fermion charge (\ref{FC}) has no
local representation. 

\section{ \mathversion{bold} Tr$\,{Q}_{\,1}$ as  an index of the Dirac
operator}

Here is the central point: the index defined in Eq. (\ref{Ind})
coincides with the square of the index of a Dirac operator on
the (reduced) moduli space of solitons,
which was studied by mathematicians.
Thus, it is possible to determine in which ${\cal N}=1$ models
${\rm Ind}_{\,\cal Z}\,(Q_2/Q_1)=0$, i.e.
the multiplet shortening does {\em not} take  place (in the general 
situation).

For every critical point $A$ the Morse index of this point $\nu(A)$
is defined as the number of the negative eigenvalues in 
the matrix of the second derivatives 
\begin{equation}
H_{ab}(\phi)=\nabla_a \partial_b {\cal W}(\phi)
\end{equation}
at $\phi =A$. At the critical points 
the covariant derivative $\nabla_a$ coincides with the regular $\partial_a$.
For solitons interpolating between two critical points, $\phi =A$ at 
$z\to -\infty$
and $\phi =B$ at $z\to \infty$
one can determine the relative Morse index $\nu_{BA}$,
\begin{equation}
\nu_{BA} = \nu (B) - \nu (A)\,.
\end{equation}
This relative Morse index counts the difference between the numbers of the zero
modes  of the operators $P$ and $P^\dagger$,
\begin{equation}
\nu_{BA}={\rm ker}\left\{P\right\}- {\rm ker}\left\{P^\dagger\right\}\,,
\end{equation}
where $P$ and $P^\dagger$ are 
\begin{equation}
P_{ab}= g_{ab} D_z - H_{ab}\,,\qquad P^\dagger_{ab}= -g_{ab} D_z - H_{ab}\,.
\end{equation}
Here $D_z$ is defined in Eq.~(\ref{covd}), and the field $\phi$ is taken to be 
$\phi_{\rm sol}(z)$. 

For the BPS soliton, satisfying Eq.~(\ref{AAone}), one zero mode certainly 
present
in $P$ is the translational mode. It corresponds to the soliton center 
$z_0$, one of
the coordinates in the soliton moduli space. The same zero mode of  $P$  is the
fermion zero mode --- the corresponding modulus $\eta$ is the superpartner of
$z_0$.

We will limit ourselves to the case when $ {\rm ker}\,{P^\dagger}=0$.
(Note that even if that is not the case, one can get rid of the zero modes in
$P^\dagger$ by small deformations of the superpotential). Then, the Morse 
index 
\begin{equation}
\nu_{BA}\equiv n+1 \ge 1
\end{equation}
counts  the dimension of the soliton moduli space
${M}^{n+1}$. Thus, we arrive at quantum mechanics of $n+1$ bosonic
and $n+1$ fermionic moduli on ${M}^{n+1}$.

The simplest case $n=0$ was analyzed in \cite{1d}. In this case the
quantum moduli  dynamics is trivial, and the single state BPS multiplet does
exist, ${\rm Ind}_{\,\cal Z}\,(Q_2/Q_1)=1$.  We will see shortly that for 
odd $n$ the index ${\rm Ind}_{\,\cal Z}\,(Q_2/Q_1)=0$, and  the soliton 
supermultiplets are
long (or reducible). We will start from a less trivial case of even $n$,
only in this category can one  expect to
find Tr$\,Q_1\neq 0$.

As was mentioned above one of $n+1$ bosonic
 moduli is $z_0$, the  coordinate of
the soliton center.
This is a cyclic coordinate conjugated to 
the generator $P_z$ of the spatial translations, $z_0 \in  {\cal R}$.
Note an ambiguity in $z_0$ --- one can add to $z_0$ an arbitrary function of 
other moduli. This ambiguity is fixed by the definition given below, see
Eq.~(\ref{mzero}).  Thus, 
the moduli space ${M}^{n+1}$ is a direct product
\begin{equation}
{M}^{n+1}={\cal R}\otimes {\cal M}^{n}
\label{RM}
\end{equation}
of ${\cal R}$ and the manifold ${\cal M}^{n}$ with coordinates  $m^1,...,m^{n}$
describing internal degrees of freedom of the soliton.  This manifold ${\cal
M}^{n}$  can be called the reduced moduli space. 

It is instructive to elucidate the
factorization (\ref{RM}) in more detail.
We must show that the moduli space metric
$h_{ij}$,
\begin{equation}
h_{ij}(m) =\!\int\! {\rm d}z\, g_{ab}(\phi_{\rm sol})\,\,
\frac{\partial \phi_{\rm sol}^a}{\partial m^i}\,\,
\frac{\partial \phi_{\rm sol}^b}{\partial m^j}\,,\qquad i,j=0,1,2,...,n\,,
\label{Aone}
\end{equation}
where $m^0\equiv z_0$, has a block form,
i.e. $h_{0j}=0$ for $j=1,2,...,n$.
Indeed,
\begin{equation}
h_{0j}(m)\, =- \!\int\! {\rm d}z\,\partial_b {\cal W}(\phi_{\rm sol}) \, 
\frac{\partial\phi_{\rm sol}^b}{\partial m^j}
 \, \,
= - \frac{\partial}{\partial m^j}\!\int\!{\rm d}z\left[ \,{\cal
W}(\phi_{\rm sol})-{\cal W}(\phi_{\rm
sol})_{m=m_*}\right]\,,
\label{A2}
\end{equation}
where we use the fact that the soliton solution depends on
the spatial coordinate  only through 
the combination $z-z_0$, to replace
${\partial \phi_{\rm sol}^a}/{\partial m^0}$
by ${\partial \phi_{\rm sol}^a}/{\partial z}$,
which, in turn, can be replaced by $\, g^{ac}\,\partial_c \,{\cal
W}(\phi_{\rm sol})$ by virtue of Eq.~(\ref{AAone}).
We also regularized 
the integral on the right-hand side
of Eq. (\ref{A2})  by subtracting from the integrand
the superpotential at some fixed values of the moduli 
$m = m_*$. 

Considering Eq.~(\ref{A2}) for $h_{00}$ we get 
\begin{equation}
h_{00}= - \frac{\partial}{\partial m^0}\!\int\!{\rm d}z\left[ \,{\cal
W}(\phi_{\rm sol})-{\cal W}(\phi_{\rm
sol})_{m=m_*}\right]\,.
\end{equation}
Having in mind $h_{00}={\cal Z}$ we define the modulus $m^0$ as  
\begin{equation}
m^0= -\frac{1}{{\cal Z}} \!\int\!{\rm d}z\left[ \,{\cal
W}(\phi_{\rm sol})-{\cal W}(\phi_{\rm
sol})_{m=m_*}\right]\,.
\label{mzero}
\end{equation}
With this definition it is clear that  
\begin{equation}
h_{0j} ={\cal Z}\,\frac{\partial m_0}{\partial m^j}=0\,,\quad (j=1,...,n)\,.
\end{equation}

Thus, the Lagrangian 
describing the moduli dynamics has the form
\begin{equation}
L\left({ M}^{n +1}\,\right)=-{\cal Z}
+\frac{{\cal Z}}{2}\left[(\dot
z_0)^2 +i
\,\eta\,\dot\eta\,
\right] +
L\left({\cal M}^{n }\right)\,,
\label{Athree}
\end{equation}
where $L\left({\cal M}^{n }\right)$
is the Lagrangian of the internal moduli, both bosonic and fermionic,
a sigma-model quantum mechanics on ${\cal M}^{n }$. We see, that the motion of
the center of mass (together with its fermionic partner) is factored out, 
and we only need to consider the dynamics on ${\cal M}^{n}$.

Quantization of $L\left({\cal M}^{n}\right)$
is standard. All operators
act in 
the Hilbert space of the spinor wave functions $\Psi_\alpha (m)$, 
where $\alpha=1,\ldots, 2^{n/2}$. The operators $m^i$ act as multiplication,
while $\dot m_i$ become matrix-differential operators.
The fermion moduli (their anticommutators form a
Clifford algebra)
become $\gamma$ matrices of dimension
$2^{n/2}\times 2^{n/2}$. 
Remember, $n$ is even, so,  there is
$\gamma^{n+1}=\prod_{i=1}^{i=n} \gamma^i$,  an analog of $\gamma^5$ 
in four dimensions. On the moduli space ${\cal M}^{n}$ 
the supercharges (\ref{supercur}) take the form
\begin{equation}
Q_1=\sqrt{2{\cal Z}}\, \gamma^{n+1}\,,\qquad
 Q_2=-\frac{i}{\sqrt{2}}\;\gamma^j\,\nabla_j \,,
\label{qq2}
\end{equation}
where the covariant derivative $\nabla_j$ includes spin connection
(for more details see \cite{LSV1}).
The expression for $Q_2$ is in fact the Dirac operator $i\!\!\not \!
\nabla$ on
${\cal M}^n$. Moreover, the Hamiltonian takes the form,
\begin{equation}
H-{\cal Z}=Q_2^2=\frac{1}{2}\,(i\!\!\not \!\nabla)^2=
-\frac{1}{2}\,\nabla^j\nabla_j+
\frac{1}{8}\,\tilde R\,,
\label{lich}
\end{equation}
 where we used the famous Lichnerowicz formula, and $\tilde R$ is the 
curvature in the soliton moduli space.

From Eqs.~(\ref{qq2}), (\ref{lich}) it is clear that the BPS soliton states 
are in correspondence with the zero modes of the Dirac operator
$i\!\!\not \! \nabla$ on
${\cal M}^{n }$.
 The index ${\rm Ind}_{\,\cal
Z}\,(Q_2/Q_1)$ we defined in Eq.~(\ref{Ind}) becomes the square of the index 
of the
Dirac operator
\begin{eqnarray}
&&{\rm Ind}_{\,\cal Z}\,(Q_2/Q_1)=\left\{{\rm Ind}\,(i\!\!\not \!\nabla)_{{\cal
M}^{n}}\right\}^2, \nonumber\\[2mm]
&&
{\rm Ind}\,(i\!\!\not \! \nabla)_{{\cal M}^{n}}={\rm Tr}
\left[\,\gamma^{n+1}
\exp\left(\beta\!\!\not \! \nabla^2
\right)\right]_{{\cal M}^{n}}.
\end{eqnarray}

Equation (\ref{lich}) shows that if the curvature $\tilde R$ is positive 
everywhere 
on the soliton moduli space the Dirac operator has no zero modes, its index
vanishes, and so does the index ${\rm Ind}_{\,\cal Z}\,(Q_2/Q_1)$. Thus, 
there is no BPS solitons in this case. An explicit example \cite{LSV1}  
is provided by a sigma model on 
$S^3$. For a certain choice of the superpotential the soliton moduli 
space is the sphere $S^2$. Moreover, there exists 
 a general mathematical assertion 
\cite{Pushkar}:  for any compact ${\cal M}^{n}$ with $n\ge 2$
the index of the Dirac operator vanishes.  The proof due
to P.~Pushkar' is outlined  in 
Appendix  of \cite{LSV1}. 

Thus, for $n\ge 2$ all soliton multiplets 
are long. If, for 
accidental or other reasons, they are still BPS saturated,  they form a 
reducible representation. For example, in \ntwo models the index  
${\rm Ind}_{\,\cal Z}\,(Q_2/Q_1)$ vanishes while the BPS states do exist. 
From the standpoint of \none
they form a reducible representation for which $(-1)^F$ is well defined.

Summarizing the case of even $n$, we conclude that  only for $n=0$, when 
${\cal M}^{n}$ reduces to a point,  the index
${\rm Ind}_{\,\cal Z}\,(Q_2/Q_1)=1$, and a single-state multiplet exists.

There is no general statement for  noncompact ${\cal M}^{n}$.
Noncompact geometry of ${\cal M}^{n}$  may emerge if there is an 
extra critical point $C$ such that
${\cal Z}_{AB}={\cal Z}_{AC}+{\cal Z}_{CB}$. 
Physically it means that the soliton 
${AB}$ is a threshold bound state of lighter solitons ${AC}$ and ${CB}$.

Return  now to the case of odd $n$.  Now, there is no matrix
$\gamma^{n+1}$ on ${\cal M}^{n}$;
the number of spinorial components of the  wave functions jumps by a
factor of two compared to the previous even value of $n$, becoming $2\times
2^{(n-1)/2}$.  In this case  the
realization of the fermion moduli can be chosen as follows:
\begin{eqnarray}
\eta=\frac{1}{\sqrt{2}}\,I\times \sigma_3\,,\qquad
\eta^i=\frac{1}{\sqrt{2}}\,\gamma^i\times \sigma_1\,,\quad (i=1,\ldots,n)\,,
\label{neven}
\end{eqnarray}
where the matrices $\gamma^i$ are are of the same dimension as in the previous
even $n$. Since $Q_1\propto \eta$,  the representation  (\ref{neven}) clearly 
demonstrates  that ${\rm Tr}\, Q_1=0$.

\section{Outlook}

The first example of the single-state supersymmetric multiplet was suggested
by Witten\cite{Ed} in the context of 2+1 supergravity with the conic geometry.
 This example was thoroughly studied in Refs.~\cite{BeckStr, JoseFidel} where
the BPS  solitons were explicitly constructed. Out of four supercharges of the
model two supercharges annihilate the BPS solitons. The other two supercharges
produce the fermion zero modes. Without gravity these modes are normalizable
which leads to two-state supermultiplet. With gravity switched on the fermion
modes become non-normalizable, implying the single-state supermultiplet.
This means that in the physical sector of the localized states all 
supercharges 
act on the soliton trivially.

In our \none examples of the single-state supermultiplet one of two 
supercharges is realized nontrivially, $Q_1=\pm\sqrt{2{\cal Z}}$.  In terms 
of modes there is one
normalizable fermion  mode. To compare with Witten's example it is convenient
to have in mind an infrared regularization --- placing the system in a 
finite  spatial 
box with supersymmetric boundary conditions. Then the normalizability is not a
criterion, and the number of the fermion modes is always even. One can trace,
however, the localization of the modes. In Witten's case both zero modes are
localized at the boundary. In our case one mode is localized on the soliton, 
while
the other at the boundary, see Ritz {\em et al.} in \cite{1d}. If one 
considers the
entire system, including the boundary, the fermion parity $(-1)^F$ is 
defined. It is
not defined, however, for localized states far away from the boundary. Similar
run-away behavior of the modes occurs in fractional charge and other phenomena
known in solid state physics.

A different way to maintain  the fermion parity $(-1)^F$ in the soliton sector
was suggested by Zamolodchikov \cite{Zamol}.
He studied a certain massive perturbation of the tricritical Ising model (in the
field theory limit it leads to supersymmetric theory with the cubic
superpotential). Rewriting the model in new variables Zamolodchikov 
arrives at three vacua instead of two.  As a result, he doubles the multiplicity of
the solitons, thus maintaining $(-1)^F$. Algebraically, it is a reducible
representation of supersymmetry.  From our point of view based on the
quasiclassical quantization this doubling is unphysical (for localized particles)
and should be eliminated. One of the possible ways to do it is through 
``orbifoldization'' (see Ref.~\cite{FendleyGinsparg} for the
relevant discussion in terms of lattice variables).   Zamolodchikov's idea was
applied recently \cite{FendleySaleur} to  the Gross-Neveu model (which for
$N=3$ is equivalent to the supersymmetric sine-Gordon model). 

Here we would like to stress a general nature of the above delocalization
phenomenon which  refers equally to minimal and extended supersymmetries.
In any theory which is completely regularized in the infrared, the BPS 
shortening,
strictly speaking, does not take place. Extra states living on the boundary 
make multiplets long, the BPS shortening is only  recovering in the 
field-theoretic limit of the infinite volume. 

In conclusion let us summarize our main points. We considered the previously
established phenomenon of the multiplet shortening in a more general class of
\mbox{\none} models in two dimensions. The generalization consists of
introducing a non-flat met\-ric on the target space. In most cases the 
shortening does not take place. In those rare cases when it does, it comes 
at a price of loosing the  fermion parity $(-1)^F$, i.e. the disappearance 
of the boson-fermion classification.

To count such short multiplets we introduce a new index (\ref{Ind}) 
(see also (\ref{IndL})). 
This index turns out to be related to the index of the Dirac operator
on the reduced soliton moduli space. The latter vanishes for all compact moduli
manifolds
 implying the absence of shortening. Finally, we  generalize the anomaly in
the central charge to take into account the target space metric.

\subsection*{Acknowledgments}

We are grateful to Jose Edelstein, Paul Fendley, Leonid Glazman, Fred Goldhaber,
Anatoly  Larkin, Bob Laughlin, Rafael Nepomechie, Peter van Nieuwenhuizen,
Pyotr Pushkar',  Adam Ritz, Andrei Smilga, Boris Shklovsky, Alexei Tsvelik, 
Misha Voloshin, Ed Witten, and Sasha Zamolodchikov for  valuable discussions.

The work of A.L. was supported in part by RFFI grant 00-02-16530,
Support for Scientific Schools grant 00-15-96-557,  and INTAS grant
99-590, the work of M.S. and A.V. was supported in part by  DOE under  grant 
DE-FG02-94ER408.

\end{document}